\documentclass[letterpaper]{article}
\usepackage{preprint}

\usepackage[hyphens]{url}
\usepackage{graphicx}
\graphicspath{{figures/}}
\usepackage{natbib}
\usepackage{amsmath}
\usepackage{booktabs}
\usepackage{multirow}
\newcommand{\method}{\textsc{RTLCurator}}

\title{RTLCurator: Label-Efficient Data Curation for RTL Generation}
\author{
Siyang Cai\textsuperscript{1,3},
Cangyuan Li\textsuperscript{3},
Wenjing Chang\textsuperscript{4},
Kun Wang\textsuperscript{2,3},\\
Haoyu Gao\textsuperscript{1,3},
Yinhe Han\textsuperscript{3},
Ying Wang\textsuperscript{3}
}
\affiliations{
\textsuperscript{1}School of Advanced Interdisciplinary Sciences, University of Chinese Academy of Sciences, Beijing, China\\
\textsuperscript{2}Hangzhou Institute for Advanced Study, University of Chinese Academy of Sciences, Hangzhou, China\\
\textsuperscript{3}CICS, Institute of Computing Technology, Chinese Academy of Sciences, Beijing, China\\
\textsuperscript{4}Computer Network Information Center, Chinese Academy of Sciences, Beijing, China\\[0.3em]
\small\mbox{caisiyang23@mails.ucas.ac.cn\quad wangying2009@ict.ac.cn}
}
\date{}

\begin{document}
\maketitle
\vspace*{-1.6em}

\begin{abstract}
Training large language models (LLMs) to write register-transfer level (RTL) requires large corpora of paired specifications and code, and such data is scarce enough that most public corpora are now synthesized. Synthesis provides scale but not correctness, and in two widely used RTL datasets only 24.4\% and 53.5\% of pairs pass generated functional tests. This raises the question of how much of such a corpus to keep and which part of it. Correctness alone is a poor answer. A pair that misbehaves in one corner case still shows valid syntax and interface conventions, and complex sequential designs are both harder to generate and harder to validate, so filtering by correctness leaves a corpus of short and simple modules. Correctness is also hard to obtain, since behavior leaves little trace on the surface in RTL, and validating an entire corpus only sorts pairs into passed and failed. We present \method{}, which learns a behavior-aware compatibility prior by contrasting each specification with implementations that fail simulation, and calibrates it to a new corpus using a small number of validated pairs. It then constructs the retained subset by balancing alignment, representation coverage, and RTL structural richness. On CodeV and RTLCoder, keeping 80\% of the corpus this way improves on training with the full corpus across all reported metrics while validating only 10\% of the pool, whereas ranking by the score alone falls below random selection and filtering the whole pool by simulation does no better. 

\end{abstract}
\vspace*{-0.4em}

\section{Introduction}

Modern chip design turns system-level requirements into physical silicon through a long and highly specialized design flow. Once architectural and microarchitectural decisions have been made, designers must encode the intended hardware behavior in RTL code, establishing the pivotal implementation-level representation that connects high-level design intent to logic synthesis and subsequent physical design. 
Generally, producing high-quality RTL requires substantial expert effort to faithfully capture hardware behavior in a form that synthesizes into an efficient implementation, and remains a major human bottleneck in the flow.

\begin{figure}[t]
    \centering
    \includegraphics[width=\columnwidth]{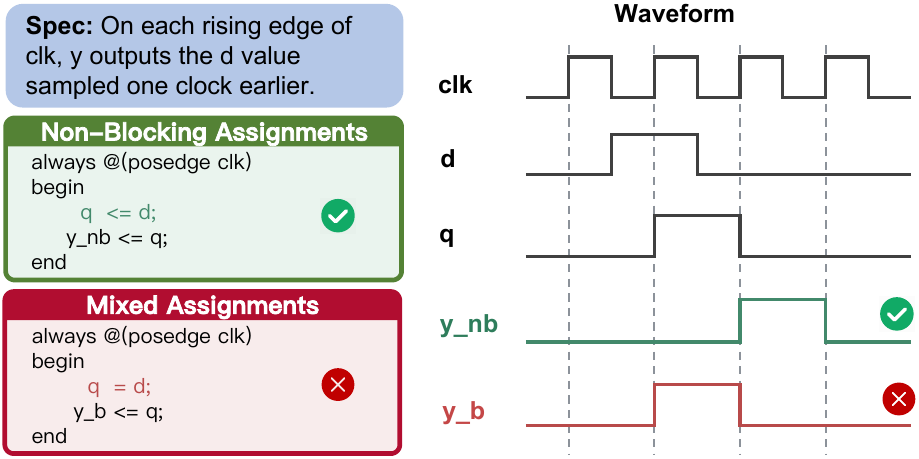}
    \caption{A one-operator edit can preserve syntax and structure while violating cycle-level spec--RTL alignment.}
    \label{fig:intro}
\end{figure}

Large language models offer a way to automate this step, and recent domain-adapted models already show measurable gains on hardware code generation \citep{RTLCoder2024,OriGen2024,CodeV2024}. Training such a model requires a large number of aligned specification--RTL (spec--RTL) pairs, and this data is hard to obtain. High-quality RTL is tied to protected intellectual property (IP) and internal design flows, and even the code that is public rarely comes with the natural language specification that would make it a usable training target. Publicly available Verilog code resources are over two orders of magnitude scarcer than resources for general-purpose programming languages
\citep{DataAllYouNeed2024}, so collecting real data at the scale these models need is not realistic.

Recent work builds training corpora by synthesis. Instructions and responses are generated with a strong model and filtered by a syntax checker \citep{RTLCoder2024,DataAllYouNeed2024}, or descriptions are summarized from crawled RTL \citep{CodeV2024}. This produces corpora of the required size, but nothing in the process checks that a response actually implements what its specification asks for. An audit of two such datasets found that among 1,000 sampled examples, only 24.4\% of RTLCoder and 53.5\% of OriGen pairs passed generated functional tests \citep{VeriCoder2025}. Training on everything therefore means training on a large amount of wrong behavior, while keeping only the pairs that pass validation means throwing most of the corpus away. 
This raises a natural question: \textit{\textbf{which part of a synthetic RTL corpus should be kept, and how much of it}}?

For an existing synthetic corpus, we argue that curation should decide which pairs to keep under a fixed retention budget rather than which pairs to remove. This raises two issues. The first issue is about the \textit{retention criterion}. The obvious criterion is whether a pair passes functional verification, but a pair that fails functionally can still be correct in other respects, such as its syntax, its interface conventions, and its coding style. Complex sequential designs also fail more often than simple ones, so a corpus filtered this way is both much smaller and shifted toward short and simple modules. The second issue is about \textit{corpus-wide quality estimation}. Surface form does not reveal behavior in RTL. One specification admits many implementations that look nothing alike yet behave identically, while replacing a non-blocking assignment with a blocking one changes cycle-level behavior without changing appearance, as shown in Figure~\ref{fig:intro}. Validation does reveal behavior but is expensive and unreliable, and even validating every pair would only sort them into passed and failed, which gives no order among the majority that fails.

To this end, we propose \method{}, a verification-efficient curation framework to address both issues. To obtain an ordering that reflects behavior without simulating every pair, we train an encoder by contrasting each specification with implementations that fail simulation, so the resulting score follows behavior rather than wording and applies to every pair of a new corpus. To make that score reliable on a corpus it was not trained on, we verify a small and actively chosen part of the corpus and resolve the undecided cases by expert adjudication, then use the results to correct the score. To keep the decision from collapsing onto the highest scores, we spread the retention budget over score strata, clusters of similar specifications, and RTL structural richness. On CodeV and RTLCoder, ranking by score alone falls below random selection and shrinks the average retained module from 37.7 to 29.2 lines, and keeping only the pairs that pass verification does no better even though it requires verifying everything. Balancing the three criteria at an 80\% budget improves on training with the full corpus across all reported metrics on VerilogEval and RTLLM while verifying only 10\% of the pool. 

We make three contributions.
\begin{itemize}
\setlength{\itemsep}{2pt}
\setlength{\parskip}{0pt}
\setlength{\topsep}{3pt}
\item We frame RTL corpus curation as deciding what to keep under a retention budget, and show that behavioral correctness serves better as a signal for this decision than as the criterion itself.
\item We train a behavior-aware alignment score that transfers to unseen corpora and is corrected from a small number of validated pairs.
\item We design a selection policy that balances this score against representation coverage and structural richness.
\end{itemize}

\begin{figure*}[t]
    \centering
    \includegraphics[width=\textwidth]{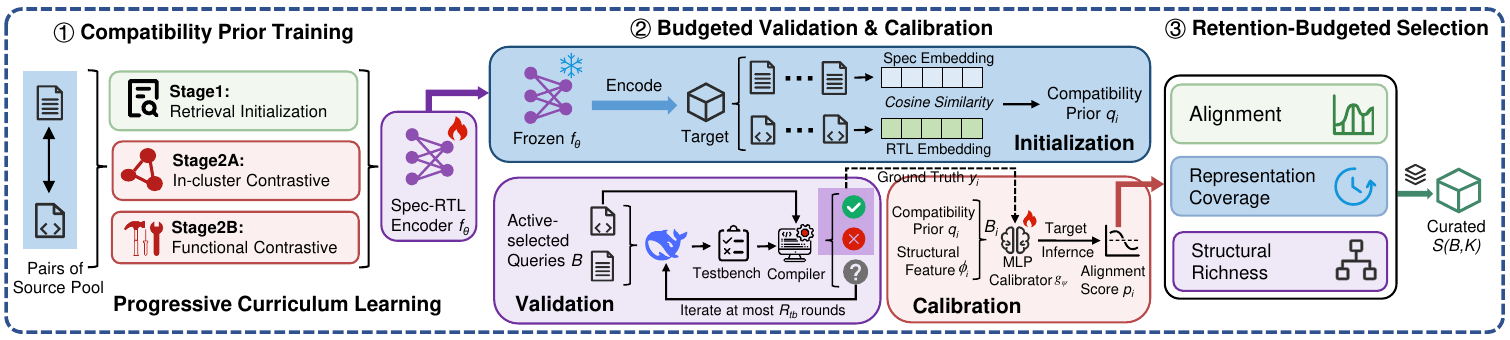}
    \caption{Overview of \method{}. An encoder is first trained on the source corpus with a progressive curriculum whose negatives move from random responses to responses that fail functional validation, which yields a compatibility prior for every pair of a target corpus before any pair is validated. A budgeted acquisition loop then queries at most $B$ pairs for validation and uses the resolved labels to calibrate this prior into an Alignment Score over the full pool. The selection stage finally distributes the retention budget $K$ across Alignment Score strata, representation clusters, and RTL structural richness.}
    \label{fig:overview}
\end{figure*}

\vspace{-0.3em}
\section{Related Work}
\vspace{-0.3em}

\paragraph{RTL corpus construction and quality control.}
Corpus construction has evolved from forward instruction--RTL synthesis with syntax filtering \citep{RTLCoder2024}, to code-grounded augmentation and reverse summarization \citep{DataAllYouNeed2024,OriGen2024,CodeV2024}, and to correct-by-construction generation \citep{CraftRTL2025}. Quality control has moved from construction-side safeguards to functional evidence. OpenLLM-RTL \citep{OpenLLMRTL2024} retains an assertion-checked subset, VeriCoder \citep{VeriCoder2025} generates unit tests and repairs pairs through simulation feedback, and CodeV-R1 \citep{CodeVR12025} and STG \citep{STG2026} screen against trusted references with equivalence checking and structured testbenches. These pipelines verify or repair every candidate they keep. We instead spend a small validation budget on an existing pool and use it to rank the rest.

\paragraph{Text--code and hardware representation learning.}
General text embeddings \citep{BGEM32025} and software code models \citep{CodeBERT2020,CodeT52021,BGECode2025} align natural language with programs for retrieval and understanding. Hardware additionally depends on structural and temporal behavior, which circuit encoders \citep{DeepGate22023,CircuitFusion2025,MGVGA2025} and RTL-specific models \citep{DeepRTL2025,DeepRTL22025,TopoRTL2026} target directly. These representations scale to a whole corpus, but similarity alone does not establish functional correctness, which is why our encoder is trained against implementations that fail simulation.

\paragraph{Data selection and label-efficient quality control.}
Code pipelines combine rule-based filters, model-based scores, and executable feedback \citep{TheStack2022,SeedCoder2025,UnitCoder2025,KodCode2025}, and instruction-data selection balances quality, diversity, influence, and rating reliability \citep{DEITA2024,LESS2024,DS2DataSelection2025}. Large-scale studies caution that score-driven selection gives limited gains once diversity collapses \citep{LargeScaleDataSelectionIT2025,RethinkingDataSelectionScale2025}, and active learning offers budgeted alternatives based on coverage, uncertainty, and gradient diversity \citep{ActiveCodeLearning2023,CoreSetAL2018,BADGE2020,MetaQueryNet2022,DeepSAD2020}. None of these signals reports whether a specification and its RTL agree, and obtaining that judgment is expensive, so we separate where validation is spent from how the retained subset is built.

\vspace{-0.3em}
\section{Method}
\vspace{-0.3em}
In this section, we present \method{}, a framework that composes an RTL training corpus under a validation budget $B$ and a retention budget $K$. The key insight is that a compatibility prior trained on functionally refuted negatives carries enough behavioral signal that a small number of validated labels suffices to calibrate it into a corpus-wide ordering. 
The entire framework is illustrated in Figure~\ref{fig:overview}.

\subsection{Problem Formulation}
\paragraph{Notation.}
Let $\mathcal{U}=\{x_i=(s_i,r_i)\}_{i=1}^{N}$ denote a target corpus of $N$ spec--RTL pairs, where $s_i$ is a specification and $r_i$ is its corresponding RTL response. For each queried pair $x_i$, the validation pipeline produces a resolved binary alignment label $y_i\in\{0,1\}$, where $1$ denotes alignment and $0$ denotes misalignment. We use $\mathcal{Q}\subseteq\mathcal{U}$ for the queried set and allow at most $B\ll N$ queried labels. We use $\mathcal{S}\subseteq\mathcal{U}$ for the retained subset and require $|\mathcal{S}|=K<N$. Here, $B$ and $K$ are the validation and retention budgets, respectively.

\paragraph{Problem Definition.}
Our goal is to use at most $B$ resolved alignment labels to select $K$ original pairs whose supervised fine-tuning yields high downstream utility.
Under a fixed model and training recipe, the ideal retained subset is
\begin{equation}
\mathcal{S}_{K}^{\star}
=
\arg\max_{\substack{\mathcal{S}\subseteq\mathcal{U}\\|\mathcal{S}|=K}}
\mathcal{J}\!\left(\mathcal{A}(\theta_0,\mathcal{S})\right),
\label{eq:ideal_curation_objective}
\end{equation}
where $\mathcal{A}$ denotes supervised fine-tuning from the initial model $\theta_0$, and $\mathcal{J}$ denotes downstream utility. The validation
budget $B$ limits the supervision available for estimating this subset, whereas $K$ determines its final size. Because downstream utility cannot be
evaluated repeatedly during curation, \method{} does not optimize Equation~\ref{eq:ideal_curation_objective} directly. 

\subsection{Training the Compatibility Prior}
We aim to learn a prior that separates implementations that are behaviorally distinct but lexically close, a distinction that random negatives leave unlearned. We initialize the encoder from the Qwen3-Embedding-4B backbone\citep{Qwen3Embedding2025} and train it in three cumulative stages of increasing negative difficulty. In the source corpus $\mathcal{D}_{\mathrm{src}}$, each specification is paired with a validated response and candidates refuted by source-side simulation.

\paragraph{Shift-Aware Preprocessing.}
Target corpora differ from $\mathcal{D}_{\mathrm{src}}$ in specification style and level of detail, shifting the input distribution seen by the source-trained encoder. To construct an encoder-familiar view, we provide both the original target specification \(s_i\) and its candidate RTL response \(r_i\) to the rewriting model and ask it to generate a specification in the source-corpus style,
\begin{equation}
\widetilde{s}_i=\mathcal{R}(s_i,r_i),
\label{eq:aux_spec}
\end{equation}
where $\widetilde{s}_i$ is an auxiliary specification used only by the encoder pathway, which
supplies the prior and the embedding geometry for clustering, acquisition routing, calibration, and coverage allocation. It does not replace $s_i$ as the behavioral contract. 
Testbench generation and validation always use the original \(s_i\), and every retained index is mapped to its original pair \((s_i,r_i)\) for LLM training.

\paragraph{Stage 1: Retrieval Initialization.}
Verified pairs and random cross-pair RTL negatives first establish broad correspondence between design intent and implementation through a multiple-negatives objective $\mathcal{L}_{\mathrm{retr}}$. This warm start creates meaningful global neighborhoods before they are used for mining, acquisition, and coverage. Random negatives alone remain insufficient because coarse functional or interface differences can separate them without resolving nearby competing implementations.

\paragraph{Stage 2A: In-Cluster Contrastive Learning.}
Using the Stage~1 embedding space, we next contrast each verified response with RTL drawn from the same representation cluster but paired with a different specification. These neighbors share broad semantic or structural context, so the encoder must decide which response actually matches the specification rather than rely on topical similarity. Replay of Stage~1 relations preserves the global organization while this local discrimination is learned.

\paragraph{Stage 2B: Functional Contrastive Learning.}
The final stage contrasts a validated response with same-specification candidates from multiple generation sources that appear plausible in the preceding embedding space but fail source-side functional validation. Holding the specification fixed removes cross-specification shortcuts, while multiple sources broaden the observed failure modes. This supervision sharpens the prior exactly where candidates are most similar, among RTL responses written for the same specification.

\paragraph{Curriculum Objective.}
All curriculum relations are formed from source-corpus records.
The encoder $f_\theta$ produces normalized specification and RTL embeddings, whose cosine similarity defines the compatibility prior
$q_\theta(s,r)=\operatorname{cos}(f_\theta(s),f_\theta(r))$.
Stages~2A and~2B share a single loss form and differ only in how the negative is drawn. For a positive response $r_i^{+}$ and a negative $r_i^{-}$, we form the margin-violation term
\begin{equation}
\Delta_i=q_\theta(s_i,r_i^{-})-q_\theta(s_i,r_i^{+})+\mu,
\label{eq:margin}
\end{equation}
where $\mu>0$ is the margin, and we take the logistic loss
\begin{equation}
\ell_i=\log\!\left(1+\exp(\Delta_i)\right).
\label{eq:hard_negative_loss}
\end{equation}
Drawing $r_i^{-}$ as an in-cluster cross-specification negative and averaging $\ell_i$ over those
relations gives $\mathcal{L}_{\mathrm{cross}}$ for Stage~2A, while drawing it as a
same-specification functional failure gives $\mathcal{L}_{\mathrm{func}}$ for Stage~2B.

The cumulative stage objectives are
\begin{equation}
\begin{aligned}
\mathcal{L}_{1}  &= \mathcal{L}_{\mathrm{retr}}, \\
\mathcal{L}_{2A} &= \lambda_{r}^{A}\mathcal{L}_{\mathrm{retr}}^{\mathrm{replay}}
                  +\ \lambda_{c}^{A}\mathcal{L}_{\mathrm{cross}}, \\
\mathcal{L}_{2B} &= \lambda_{r}^{B}\mathcal{L}_{\mathrm{retr}}^{\mathrm{replay}}
                  +\ \lambda_{c}^{B}\mathcal{L}_{\mathrm{cross}}
                  +\ \lambda_{f}^{B}\mathcal{L}_{\mathrm{func}}.
\end{aligned}
\label{eq:progressive_curriculum}
\end{equation}
where $\mathcal{L}_{\mathrm{retr}}^{\mathrm{replay}}$ replays Stage~1 relations under the current
encoder and $\lambda_{r}^{\bullet},\lambda_{c}^{\bullet},\lambda_{f}^{B}\!\ge\!0$ are stage weights.
The cumulative curriculum preserves the corpus-scale embedding space needed for scoring and coverage while sharpening sensitivity to local, behaviorally consequential mismatches.

\subsection{Budgeted Validation and Calibration}

With the source encoder fixed, \method{} assigns every target record the
compatibility prior $q_i=q_\theta(\mathcal{R}(s_i,r_i),r_i)$. It partitions the embeddings of $\widetilde{s}_i$ once into $C$ fixed clusters, where
$c(i)\in\{1,\ldots,C\}$ denotes the cluster assignment of record $i$.
The same assignments are used across all acquisition rounds and retention
budgets. Before any target label is available, routing uses the rank-normalized
compatibility prior $q_i$. After calibration begins, it uses the current
Alignment Score $p_i$.

\paragraph{Acquisition Routing.}
Each acquisition batch divides its quota evenly among five routes that target the decision boundary induced by the requested retention budget $K$, the ambiguous center of the ranking, clusters with little resolved evidence, and the two ranking tails. When several retention budgets are requested, the routes cover their associated boundaries and the acquired labels are shared across the resulting subsets. Unfilled quota is reassigned to the remaining eligible candidates. These routes determine where evidence is acquired, rather than which pairs are retained.

\paragraph{Validation Protocol.}
For each queried record $x_i$, the validation pipeline evaluates $r_i$ using compilation, simulation, and testbench evidence. When the testbench does not exercise behavior sufficient to distinguish alignment from misalignment, it is revised and rerun for at most $R_{\mathrm{tb}}$ rounds. A case that remains unresolved after this limit is escalated to a panel of LLM judges or expert review. This process produces a binary alignment label $y_i$. Generated testbenches provide evidence only, without replacing the original RTL response.

\paragraph{Target Calibration.}

Let $\phi_i$ denote the lightweight calibration features for candidate $x_i$, and let $\eta_i>0$ be its evidence-confidence weight. At round $t$, let $\mathcal{Q}_t\subseteq\mathcal{Q}$ contain the queried records resolved so far. The calibrator $g_\psi$ predicts
\begin{equation}
\widehat{p}_i^{\,\psi}=g_\psi(q_i,\phi_i),
\label{eq:calibrator}
\end{equation}
and is fit using the binary cross-entropy (BCE) loss:
\begin{equation}
\psi^\star=\arg\min_\psi\sum_{x_j\in\mathcal{Q}_t}
\eta_j\,\ell_{\mathrm{BCE}}\!\left(y_j,\widehat{p}_j^{\,\psi}\right),
\label{eq:target_calibration}
\end{equation}
The Alignment Score is
$p_i=\widehat{p}_i^{\,\psi^\star}\in[0,1]$. The calibrator is refit after each round; the updated scores route the next round, and the final fit scores all of $\mathcal{U}$ after at most $B$ labels. 

\subsection{Retention-Budgeted Selection Policy}

\method{} balances Alignment Score, Representation Coverage, and Structural Richness through a staged selection policy. Alignment Score strata determine
how the retention budget is distributed across quality regions. Representation Coverage distributes each stratum quota across clusters. Alignment Score and Structural Richness then jointly determine which records
are retained within each resulting cell.

Let $K_m$ denote the quota assigned to stratum $m$, with $\sum_m K_m=K$. The stratum with the highest Alignment Scores receives the largest nominal share. The remaining strata receive equal smaller shares. If a stratum cannot fill its quota, the excess is redistributed according to the remaining capacities. Within each stratum, $K_m$ is distributed across the fixed clusters in proportion to their candidate counts. Candidates within each resulting cell are then ranked using the Alignment Score and Structural Richness.

Let $\boldsymbol{z}_i\in[0,1]^d$ contain the normalized RTL structural features. Token, line, and port counts use percentile normalization over the full pool. Counts of control logic are clipped, while finite-state machine (FSM) and clock cues are binary. Structural Richness is defined by the fixed weighted combination
\begin{equation}
\rho(r_i)=\boldsymbol{w}^{\top}\boldsymbol{z}_i,
\qquad
\boldsymbol{w}\succeq\boldsymbol{0},
\quad
\boldsymbol{1}^{\top}\boldsymbol{w}=1.
\label{eq:structural_richness}
\end{equation}
Within each cell defined by one stratum and one cluster, a fixed bonus
determined from the quantile of $\rho(r_i)$ is added to the Alignment Score
$p_i$. 

\subsection{Complexity Analysis}

Curation costs one encoder pass and one clustering over $\mathcal{U}$, $O(N)$ and
$O(N|\mathcal{C}|)$ respectively, $T$ calibrator refits on at most $B$ labels, and one
$O(N\log N)$ sort for stratification. The dominant real cost is the $B$ calls to $\mathcal{O}$,
which is why $B\ll N$ is the budget that matters rather than $N$ itself.

\vspace{-0.3em}
\section{Experiments}
\vspace{-0.3em}
\subsection{Experimental Setup}
  \begin{table*}[!t]
\centering
\footnotesize
\setlength{\belowcaptionskip}{4pt}
\caption{Downstream supervised fine-tuning performance at the 80\% budget. \textbf{Bold}, \textit{italics}, and \underline{underline} mark the first-, second-, and third-best results within each corpus block, respectively, and ties share the same mark. Results are five-run means (\%).}
\label{tab:main_downstream}
\small
\setlength{\tabcolsep}{2.2pt}
\renewcommand{\arraystretch}{1.05}
\resizebox{0.88\textwidth}{!}{%
\begin{tabular}{llc*{10}{c}}
\toprule
\multirow{3}{*}{\textbf{Corpus}} & \multirow{3}{*}{\textbf{Policy}} & \multirow{3}{*}{\textbf{Budget}}
& \multicolumn{3}{c}{\textbf{VerilogEval Machine}}
& \multicolumn{3}{c}{\textbf{VerilogEval Human}}
& \multicolumn{2}{c}{\textbf{RTLLM v1.0}}
& \multicolumn{2}{c}{\textbf{RTLLM v1.1}} \\
\cmidrule(lr){4-6}\cmidrule(lr){7-9}\cmidrule(lr){10-11}\cmidrule(lr){12-13}
& & & \multicolumn{3}{c}{\textbf{Functional}}
& \multicolumn{3}{c}{\textbf{Functional}}
& \textbf{Syntax} & \textbf{Functional}
& \textbf{Syntax} & \textbf{Functional} \\
\cmidrule(lr){4-6}\cmidrule(lr){7-9}\cmidrule(lr){10-10}\cmidrule(lr){11-11}\cmidrule(lr){12-12}\cmidrule(lr){13-13}
& & & \textbf{@1} & \textbf{@5} & \textbf{@10}
& \textbf{@1} & \textbf{@5} & \textbf{@10}
& \textbf{@5} & \textbf{@5} & \textbf{@5} & \textbf{@5} \\
\midrule
-- & Base Model & 0\% & 56.85 & 65.23 & 67.16 & 30.80 & 41.15 & 44.43 & 60.60 & 30.05 & 65.89 & 35.18 \\
\midrule
\multirow{6}{*}{CodeV}
& Full & 100\% & \textit{77.90} & \underline{84.52} & \underline{86.52} & \underline{45.54} & 60.36 & \textit{65.72} & \underline{86.62} & 56.89 & \textit{90.72} & 52.86 \\
& Random & 80\% & \underline{76.82} & 83.77 & 85.57 & \textit{45.96} & \underline{60.41} & 64.81 & 85.94 & \underline{57.44} & 89.34 & \underline{56.29} \\
& DS$^2$ & 80\% & 75.63 & 81.12 & 82.18 & 44.68 & \textit{61.12} & \underline{65.40} & 85.79 & 56.88 & \underline{90.52} & 56.21 \\
& LESS & 80\% & 76.19 & \textit{85.43} & \textit{87.84} & 44.74 & 59.29 & 63.32 & \textit{86.94} & 56.84 & 87.82 & 56.12 \\
& Align.\ top & 80\% & 74.06 & 82.76 & 86.10 & 42.60 & 56.10 & 60.44 & 82.95 & \textit{61.40} & 89.83 & \textit{61.77} \\
& \textbf{\method{}} & 80\% & \textbf{79.22} & \textbf{89.77} & \textbf{90.38} & \textbf{47.49} & \textbf{62.45} & \textbf{67.89} & \textbf{89.05} & \textbf{63.67} & \textbf{94.31} & \textbf{63.97} \\
\midrule
\multirow{6}{*}{RTLCoder}
& Full & 100\% & \textit{65.28} & 71.36 & 73.24 & \underline{39.52} & 42.77 & 43.53 & \textit{82.50} & \underline{51.56} & \textit{84.98} & \textit{51.33} \\
& Random & 80\% & \underline{64.20} & \textit{74.09} & \textbf{76.38} & \textit{39.71} & \textit{48.40} & \underline{50.77} & 78.50 & 49.68 & \underline{82.73} & 44.94 \\
& DS$^2$ & 80\% & 61.61 & 71.09 & 73.03 & 37.69 & \underline{48.02} & \textit{51.06} & 79.78 & 47.45 & 78.74 & 45.11 \\
& LESS & 80\% & 60.70 & 70.60 & 73.45 & 39.07 & 47.61 & 50.05 & \underline{80.50} & \textit{52.01} & 81.98 & \underline{47.14} \\
& Align.\ top & 80\% & 63.60 & \underline{73.39} & \textit{75.99} & \textit{39.71} & 47.87 & 50.42 & 79.81 & 50.56 & 79.01 & 39.85 \\
& \textbf{\method{}} & 80\% & \textbf{66.24} & \textbf{74.67} & \underline{75.86} & \textbf{40.42} & \textbf{48.67} & \textbf{52.60} & \textbf{83.94} & \textbf{56.00} & \textbf{86.30} & \textbf{52.18} \\
\bottomrule
\end{tabular}%
}
\end{table*}

\label{sec:setup}
\paragraph{Datasets and Benchmarks.} We curate two synthetic corpora, CodeV \citep{CodeV2024} (164{,}111 pairs) and RTLCoder \citep{RTLCoder2024} (26{,}532 pairs), retaining 50--90\% of each pool. 
The encoder that produces the compatibility prior is trained on the public VeriCoder/OriGen release \citep{VeriCoder2025,OriGen2024} with a 90/5/5 split, which contains no CodeV or RTLCoder data. From this source corpus, we construct the Stage~2B function-hard negatives using the RTL-specialized model CodeV-R1 \citep{CodeVR12025} and two general-purpose models MiniMax-M2.7 and DeepSeek-V4-Flash. Validation labels on the two target pools are collected separately (Sec.~\ref{sec:encoder_calibration}). We evaluate on VerilogEval \citep{VerilogEval2023}, whose Machine and Human splits contain HDLBits-derived module-level specifications, and on RTLLM v1.0/v1.1 \citep{RTLLM2024}, which comprises larger design-level problems. Before selection, we remove pairs that duplicate a VerilogEval or RTLLM problem.

\paragraph{Baselines.} We compare \method{} against five selection policies. \textbf{\textit{Full}} keeps the whole corpus, while other policies retain a matched fraction of it. 
\textbf{\textit{Random}} samples uniformly. 
\textbf{\textit{DS$^2$}} \citep{DS2DataSelection2025} and \textbf{\textit{LESS}} \citep{LESS2024} are data selection methods developed for general-purpose instruction tuning. \textbf{\textit{Align.\ top}} keeps the highest-scoring pairs under our alignment score, it uses the same signal as \method{} but none of the coverage or structural balancing.

\paragraph{Metrics.} We report functional pass@1/5/10 on VerilogEval Machine/Human \citep{VerilogEval2023} and syntax and functional pass@5 on RTLLM v1.0/v1.1 \citep{RTLLM2024}. Each result is averaged over five independent draws of $k$ samples per problem.
We also report diagnostics of the encoder and the calibrator. Retrieval over the full candidate pool uses recall at 1 (R@1), mean reciprocal rank at 10 (MRR@10), and normalized discounted cumulative gain at 10 (nDCG@10). Ranking one compatible implementation against plausible failures for the same specification uses pairwise accuracy, average precision (AP), area under the receiver operating characteristic curve (AUROC), and full-group R@1, which is recall at 1 within a single specification's candidate group rather than over the pool.

\paragraph{Implementation Details.} Each selected subset is used to fine-tune DeepSeek-Coder-6.7B-Instruct \citep{DeepSeekCoder2024} under identical hyperparameters across all policies and retention budgets. Selection never modifies data, so a pair is either kept as it is or dropped. Full hyperparameters, hardware, and runtime are reported in the technical appendix.
\begin{table}[t]
\centering
\footnotesize
\setlength{\belowcaptionskip}{4pt}
\caption{Mean lines of code in the retained subsets at the 80\% retention budget. Full is the uncurated corpus, which every policy draws from.}
\label{tab:structure}
\renewcommand{\arraystretch}{1.10}
\resizebox{0.89\columnwidth}{!}{%
\begin{tabular}{lcccc}
\toprule
\textbf{Corpus} & \textbf{Full} & \textbf{\textit{Random}} & \textbf{\textit{Align.\ top}} & \textbf{\method{}} \\ 
\midrule
CodeV    & 37.6 & 37.7 & 29.2 & 33.6 \\
RTLCoder & 46.7 & 40.5 & 30.6 & 36.8 \\
\bottomrule
\end{tabular}%
}
\end{table}
\subsection{Main Results}
\label{sec:main_results}
Table~\ref{tab:main_downstream} compares all policies at a retention budget of 80\%, where every subset policy keeps the same number of pairs and differs only in which ones. Sec.~\ref{sec:budget_ablation} reports the full 50--90\% range. \textbf{\method{}} outperforms \textit{\textbf{Full}} on all ten metrics in both corpora and leads every other policy except on RTLCoder Machine pass@10, where \textit{\textbf{Random}} is 0.52 points higher. The base model, hyperparameters, and subset size are identical across policies, so the difference comes from which pairs are kept rather than how many. Neither more data nor indiscriminate pruning explains it, since \textit{\textbf{Full}} loses and \textit{\textbf{Random}} trails on 19 of these 20 comparisons. Dropping 20\% of CodeV raises Machine pass@5 from 84.52 to 89.77 and RTLLM v1.1 functional pass@5 from 52.86 to 63.97.
\begin{table}[t]
\centering
\small
\setlength{\tabcolsep}{4pt}
\caption{Full-pool validation comparison on RTLCoder. Results are five-run means (\%), and \textbf{bold} marks the best value.}
\label{tab:full_validation}
\renewcommand{\arraystretch}{1.05}
\resizebox{0.89\columnwidth}{!}{%
\begin{tabular}{lccc}
\toprule
\textbf{Dataset} & \textbf{Metric} & \textbf{\textit{Full-Val}} & \textbf{\method{}} \\
\midrule
\multirow{2}{*}{VerilogEval Machine} & F@1 & 64.30 & \textbf{66.24} \\
                                     & F@5 & 74.25 & \textbf{74.67} \\ 
\addlinespace[2pt]
\multirow{2}{*}{VerilogEval Human}   & F@1 & 40.35 & \textbf{40.42} \\
                                     & F@5 & 48.03 & \textbf{48.67} \\ 
\addlinespace[2pt]
\multirow{2}{*}{RTLLM v1.0}          & F@1 & 41.40 & \textbf{43.40} \\
                                     & F@5 & 51.36 & \textbf{56.00} \\ 
\addlinespace[2pt]
\multirow{2}{*}{RTLLM v1.1}          & F@1 & 30.11 & \textbf{37.59} \\
                                     & F@5 & 44.52 & \textbf{52.18} \\
\bottomrule
\end{tabular}
}
\end{table}

\begin{figure*}[!t]
\centering 
\includegraphics[width=\textwidth]{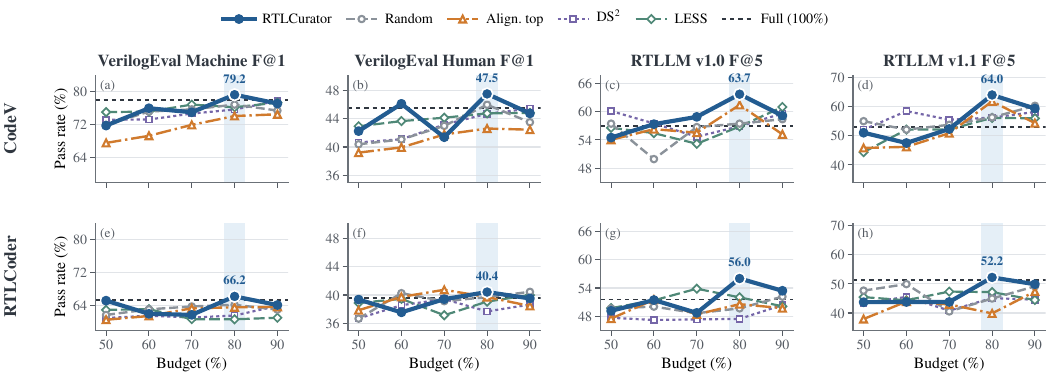}
\caption{Budget sensitivity of downstream supervised fine-tuning on CodeV and RTLCoder. Curves show five-run mean performance (\%) for all policies over 50--90\% retention, where F@\(k\) denotes functional pass@\(k\). Dotted horizontal lines denote corpus-matched Full (100\%) fine-tuning, and blue bands mark 80\%.}
\label{fig:budget_sweep}
\end{figure*}
The other policies show where the gain comes from. \textit{\textbf{DS$^2$}} and \textit{\textbf{LESS}} do not consistently improve over \textit{\textbf{Random}} and lead no cell in either corpus, falling below it on VerilogEval pass@1 in both corpora by up to 3.50 points. Neither signal reports whether a specification and its RTL implementation actually agree. \textit{\textbf{Align.\ top}} tests the opposite case. It ranks by our alignment score, which does reflect behavior, but keeps only the top of that ranking, and it also falls below \textit{\textbf{Random}} on pass@1 in both corpora. A behavior-aware score alone is therefore not enough, and it has to be spent across the corpus instead of concentrated at the top. The effect is clearest on the larger RTLLM designs, where \textit{\textbf{Random}} drops below \textit{\textbf{Full}} on RTLCoder by up to 6.39 points while \textbf{\method{}} stays above it in both corpora.

\subsection{Selection Policy Comparison}
\label{sec:policy_comparison}
Table~\ref{tab:main_downstream} also compares three policies at the same 80\% budget. \textit{\textbf{Align.\ top}} keeps the alignment ranking but drops the cluster-proportional and structure-aware allocation. \textbf{\method{}} exceeds both controls on all six metrics in both corpora, by up to 5.52 points on CodeV and 7.24 on RTLCoder over whichever control is stronger, while \textit{\textbf{Align.\ top}} does not consistently improve over \textit{\textbf{Random}}. Mean RTL length shows what goes wrong. As shown in Table~\ref{tab:structure}, random selection already shortens the average RTLCoder module from 46.7 to 40.5 lines, \textit{\textbf{Align.\ top}} shrinks it to 30.6, and \textbf{\method{}} keeps 36.8 while performing best. Ranking by score alone pulls the subset toward short, easily satisfied modules, which coverage and structural balance counteract together.

Replacing the learned score with simulation gives a stronger version of the same comparison. Using the released VeriCoder testbenches \citep{VeriCoder2025}, \textit{\textbf{Full-Val}} validates the entire RTLCoder pool and retains the passing subset, about 20\% of it, with no coverage or structural balance. Table~\ref{tab:full_validation} shows that \textbf{\method{}} outperforms it on all reported metrics while validating only 10\% of the pool. Only 19.9\% of pairs pass, so the filtered subset is four times smaller than ours and also structurally narrower, with a mean module length of 30.8 lines against 36.8.

\begin{table*}[!t]
\centering
\caption{Cross-corpus retrieval between specifications and RTL code (\%). OriGen is in-domain (ID) for the encoder, which was trained on it, whereas the two target corpora are out-of-domain (OOD). \(N\) is both the query count and the ranked RTL candidate count. `Raw' uses the original target specifications. `LLM' uses specifications regenerated in the source-corpus style from both the original specification and its candidate implementation. \textbf{Bold} and \underline{underline} denote the best and second-best values, respectively.}
\label{tab:encoder_retrieval}
\fontsize{6.8}{7.6}\selectfont
\setlength{\tabcolsep}{0.6pt}
\renewcommand{\arraystretch}{0.96}
\resizebox{\textwidth}{!}{%
\begin{tabular}{@{}l*{15}{c}@{}}
\toprule
\multirow{3}{*}{\textbf{Model}}
& \multicolumn{3}{c}{\textbf{OriGen} \textnormal{(ID)}}
& \multicolumn{6}{c}{\textbf{CodeV} \textnormal{(OOD)}}
& \multicolumn{6}{c}{\textbf{RTLCoder} \textnormal{(OOD)}} \\
\cmidrule(lr){2-4}\cmidrule(lr){5-10}\cmidrule(lr){11-16}
& \multicolumn{3}{c}{\shortstack{\textbf{Test}\\[-1pt]{\scriptsize \(N=6{,}086\)}}}
& \multicolumn{3}{c}{\shortstack{\textbf{Raw}\\[-1pt]{\scriptsize \(N=164{,}111\)}}}
& \multicolumn{3}{c}{\shortstack{\textbf{LLM}\\[-1pt]{\scriptsize \(N=164{,}111\)}}}
& \multicolumn{3}{c}{\shortstack{\textbf{Raw}\\[-1pt]{\scriptsize \(N=26{,}532\)}}}
& \multicolumn{3}{c}{\shortstack{\textbf{LLM}\\[-1pt]{\scriptsize \(N=26{,}532\)}}} \\
\cmidrule(lr){2-4}\cmidrule(lr){5-7}\cmidrule(lr){8-10}
\cmidrule(lr){11-13}\cmidrule(lr){14-16}
& \textbf{R@1} & \textbf{MRR@10} & \textbf{nDCG@10}
& \textbf{R@1} & \textbf{MRR@10} & \textbf{nDCG@10}
& \textbf{R@1} & \textbf{MRR@10} & \textbf{nDCG@10}
& \textbf{R@1} & \textbf{MRR@10} & \textbf{nDCG@10}
& \textbf{R@1} & \textbf{MRR@10} & \textbf{nDCG@10} \\
\midrule
BGE-M3
& 30.09 & 36.75 & 40.57
& 10.40 & 13.91 & 15.98
& 13.30 & 17.15 & 19.37
& 19.23 & 25.60 & 29.25
& 13.03 & 17.22 & 19.59 \\
BGE-Code-v1
& 73.92 & 79.67 & 82.39
& 11.44 & 15.45 & 17.75
& 30.79 & 37.17 & 40.63
& 27.01 & 34.03 & 37.90
& 32.01 & 39.95 & 44.25 \\
Qwen2.5-Coder-3B
& 0.67 & 1.47 & 2.06
& 0.04 & 0.08 & 0.11
& 0.08 & 0.17 & 0.23
& 0.24 & 0.47 & 0.64
& 0.17 & 0.39 & 0.53 \\
CodeV-R1-Qwen-7B
& 25.34 & 32.81 & 36.99
& 2.34 & 3.63 & 4.45
& 4.50 & 6.77 & 8.17
& 6.15 & 9.63 & 11.81
& 7.17 & 11.05 & 13.44 \\
Qwen3-Emb-8B
& 79.67 & 85.22 & 87.67
& 19.66 & 24.93 & 27.91
& 32.69 & 39.43 & 43.06
& 35.39 & 44.19 & 48.99
& 40.13 & 48.50 & 52.91 \\
Qwen3-Emb-4B
& \underline{82.93} & \underline{87.48} & \underline{89.45}
& \underline{25.03} & \underline{31.13} & \underline{34.52}
& \underline{45.44} & \underline{53.26} & \underline{57.27}
& \underline{42.06} & \underline{51.20} & \underline{56.11}
& \underline{54.76} & \underline{63.69} & \underline{68.09} \\
\midrule
\textbf{\method{}}
& \textbf{97.39} & \textbf{98.45} & \textbf{98.83}
& \textbf{36.74} & \textbf{43.11} & \textbf{46.46}
& \textbf{67.02} & \textbf{73.58} & \textbf{76.67}
& \textbf{53.46} & \textbf{60.82} & \textbf{64.43}
& \textbf{64.39} & \textbf{71.85} & \textbf{75.42} \\
\bottomrule
\end{tabular}%
}
\end{table*}

\subsection{Retention Budget Sensitivity}
\label{sec:budget_ablation}
Figure~\ref{fig:budget_sweep} sweeps the retention budget from 50\% to 90\% with each policy held fixed. Utility does not grow with the budget. \textbf{\method{}} peaks at 80\% and falls at 90\% on both corpora and all four metrics shown, the 80\% subset also outperforms \textit{\textbf{Full}}, and no other policy peaks at 80\% in both corpora. The peak sits at 80\% even though far fewer than 80\% of pairs in these corpora pass a functional test \citep{VeriCoder2025}. A pair whose behavior deviates from its specification can still show valid syntax, interface conventions, and coding style, so removing pairs aggressively loses usable supervision faster than it removes noise. What matters is which pairs are kept rather than how many, and the agreement between VerilogEval and RTLLM shows this is not specific to one task size.
\subsection{Encoder and Calibration Analysis}
  \label{sec:encoder_calibration}
\paragraph{Cross-Corpus Retrieval.} Table~\ref{tab:encoder_retrieval} asks whether a specification retrieves its own implementation out of the full pool. On held-out OriGen data the encoder reaches 97.39 R@1, so training on functional contrasts does not damage matching on the source corpus. It also transfers. Against the strongest baseline, Qwen3-Emb-4B, \textbf{\method{}} gains 11.71 and 11.40 R@1 points on the original CodeV and RTLCoder specifications, with similar margins on MRR@10 and nDCG@10, while ranking against 164{,}111 and 26{,}532 candidates respectively. The LLM column evaluates the preprocessing view, which regenerates each specification in the source-corpus style from both the original specification and its candidate implementation, and all models are given the same rewritten queries. Under that condition \textbf{\method{}} gains about 30 points on CodeV and keeps a margin of at least 19.40 points over the strongest baseline. RTLCoder gains about 11 points over its 24{,}780 rewritten records. The larger CodeV shift matches how that corpus was built, since its specifications are summaries written from crawled RTL.

\begin{figure}[t]
\centering
\vspace*{-10pt}
\includegraphics[width=\columnwidth]{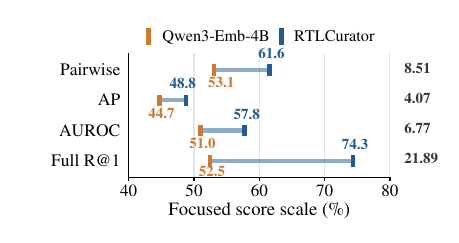}
\caption{Functional-proxy ranking on held-out OriGen test-set groups containing compatible RTL code and multi-source functional failures. Qwen3-Emb-4B abbreviates Qwen3-Embedding-4B. Endpoints report scores (\%), and right-side labels report gains in percentage points (pp).}
\label{fig:encoder_quality_proxy}
\vspace{-0.6em}
\end{figure}

\paragraph{Ranking Correct against Failing Implementations.} Retrieval does not show whether the encoder can tell a correct implementation from a wrong one written for the same specification. We therefore build a second test from the held-out OriGen data, pairing each specification with one implementation that passes its testbench and several that look reasonable but fail (4{,}181 groups, 19{,}131 candidates). As Figure~\ref{fig:encoder_quality_proxy} shows, the trained encoder improves over its initialization backbone by 8.51 points in pairwise accuracy, 4.07 in AP, 6.77 in AUROC, and 21.89 in picking the correct implementation out of its whole group. The last number is the important one, since picking the one correct implementation from a field of plausible failures is exactly the decision selection has to make.

\begin{figure}[t]
\centering
\includegraphics[width=0.91\columnwidth]{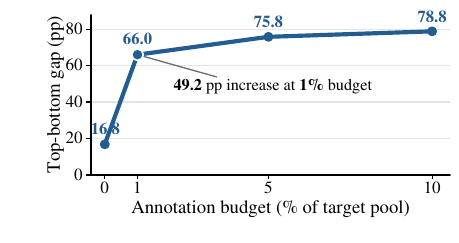}
\caption{RTLCoder target-domain calibration across cumulative validation budgets. At each budget, records are ranked by the current score, and the plotted top--bottom gap is the observed accept rate among the top 1,000 records minus that among the bottom 1,000. The \(0\%\) point uses the uncalibrated encoder prior, and pp denotes percentage points.}
\label{fig:target_calibration}
\vspace{-0.6em}
\end{figure}

\paragraph{Effect of the Validation Budget.} Figure~\ref{fig:target_calibration} tracks what target labels buy on RTLCoder. At each budget we rank the pool by the current score and measure the gap in accept rate between the top 1{,}000 and the bottom 1{,}000 records. Before any target label the gap is 16.8 points. The first 1\% of labels widens it to 66.0 points, and 5\% and 10\% reach 75.8 and 78.8, so the first 1\% accounts for about 79\% of the total widening. At the largest budget the loop has labeled 2{,}700 examples, and the top and bottom groups accept at 85.3\% and 6.5\%. Selection uses the ranking rather than a fixed cutoff, so this gap is the quantity that matters, and it is reached after labeling a small part of the pool.

\vspace{-0.25em}
\section{Conclusion}

In this work, we present \method{}, which overcomes the limitation of validation-based curation by scoring every pair for spec--RTL alignment before any pair is validated and by balancing that score against representation coverage and structural richness. This keeps a large corpus while removing the pairs that hurt most, and it shows that behavioral correctness serves better as an input to curation than as its objective. Experiments on CodeV and RTLCoder confirm that the selected subsets outperform full-corpus fine-tuning and every matched-budget baseline. The alignment score transfers to corpora it was never trained on, which suggests it can support other hardware data pipelines. Isolating the contributions of coverage and structural richness is a valuable direction for future research.

\bibliographystyle{plainnat}
\bibliography{bib/refs}
\end{document}